\documentclass[prb,aps,preprint]{revtex4}
\usepackage[T1]{fontenc}
\usepackage{lmodern}
\usepackage{graphicx}
\usepackage{wrapfig}
\usepackage{amsmath}
\usepackage{color}
\usepackage{amssymb}
\usepackage[T1]{fontenc}
\usepackage[ansinew]{inputenc}
\usepackage[english]{babel}
\usepackage[left=2.5cm,right=2.5cm,top=2cm,bottom=3.0cm,a4paper]{geometry}
\usepackage{tabularx}
\usepackage{color}
\usepackage{dcolumn}
\usepackage{bm}

\newcommand{\be}{\begin{equation}}
\newcommand{\ee}{\end{equation}}
\newcommand{\bea}{\begin{eqnarray}}
\newcommand{\eea}{\end{eqnarray}}
\newcommand{\beq}{\begin{equation}}
\newcommand{\eeq}{\end{equation}}

\begin{document}

\title{Fe-based superconductors: seven years later}
\author{Andrey V. Chubukov}

\affiliation {Department of Physics, University of Minnesota,
Minneapolis, MN 55455, USA}

\author{P.J.~Hirschfeld}
\affiliation{Department of Physics, University of Florida, Gainesville, Florida 32611, USA}

\date{\today}


\begin{abstract}
Iron-based superconductors were discovered seven years ago, in 2008. This short review
  summarizes what we learned about these materials over the last seven years, what are open questions, and what new physics we expect to extract
   from studies of this new class of high-temperature superconductors.
  \end{abstract}

\maketitle

\section{Introduction}
\label{sec:1}

The understanding of the mechanism which
binds electrons into pairs and allows then to conduct electricity without dissipation is one of the most challenging and at the same time  most exciting issues
 in the physics of correlated electrons.  In this respect, the 2008 discovery  of high-temperature superconductivity in a class of materials based on iron~\cite{bib:Hosono},
  was arguably among the most significant breakthroughs in condensed matter physics in the last two decades.
Suddenly, in addition to the famous copper-based superconductors, researchers had a new class of  materials exhibiting the macroscopic quantum phenomenon of superconductivity at high temperatures, and it looked as though the road to room-temperature superconductivity might be smoother because of the chance to compare and contrast these two systems.
   The discovery of $Fe$-based  superconductors (FeSCs) signaled, in the minds of many, the
  transition  from the  "copper age" to the new "iron age".

In "conventional" superconductors (Pb, Hg, Nb...), electrons Bose condense at temperatures of a few Kelvin after binding into so-called "Cooper pairs".  The attractive force between the two electrons in a pair is provided by the polarization of  the crystal  lattice of positively charged ions.
 The enduring fascination with the  Cu-based and now the Fe-based superconductors arises because, in addition to their high critical temperatures $T_c$, they appear to belong to an "unconventional" class of materials, in which the binding of electrons into Cooper pairs somehow occurs {\it only} via the   repulsive Coulomb interaction without significant help from the ionic lattice of ions.

The understanding of how superconductivity can possibly emerge from repulsion alone is a notoriously difficult task
   and there is still no universally acceptable scenario for superconductivity in the cuprates after nearly 30 years.  The initial hope after the discovery of FeSCs was that, since the
    screened Coulomb interaction in these materials
     is
     generally weaker than in the cuprates,  the problem might be theoretically more tractable.  It was hoped that   it would be possible to find a consensus about  the pairing mechanism in FeSCs,  then apply this knowledge to the cuprates.  This idea is still alive, but in seven years since the discovery a collective effort by condensed-matter community has
  led to understanding  that the physics of FeSCs is far richer than originally thought, and that these materials display a number of highly non-trivial properties which have no analogs in other classes of materials.
   Here we report on the significant and exciting progress made in the six years since Charles Day summarized the early experiments and theoretical works for Physics Today readers in 2009
 [Physics Today {\bf 62} (8), 36 (2009)].

\subsection{Materials} 
\begin{figure}[tbp]
\includegraphics[angle=0,width=0.8\linewidth]{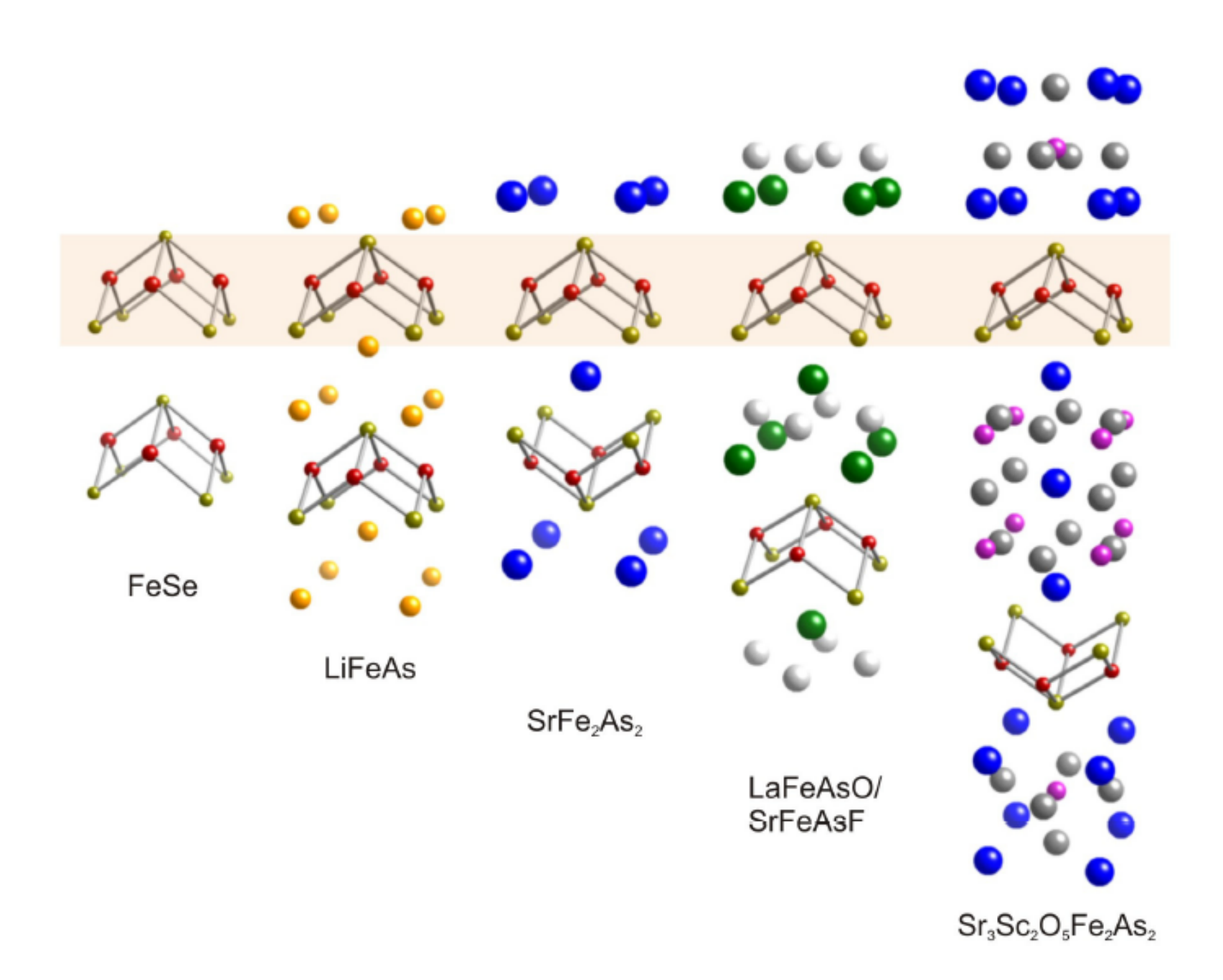}
\caption{Crystallographic structures of various families of iron-based superconductors.
 Common to all systems is the set of square lattices of Fe atoms with pnictogen or chalcogen atoms (As and Se, respectively, in the examples in the figure)
  are located above and below Fe plane, in "chess" order.
  From Ref. ~[\onlinecite{review_ex}b]. }
\label{fig:el_structure}
\end{figure}

The family of FeSCs  is already quite large and
keeps growing. It includes various Fe-pnictides and Fe-chalcogenides
 (pnictogens are elements group 15: N,P, As,Sb, Bi, and chalcogens are elements from the group 16: O, S, Se, Te).
Examples of Fe-pnictides are
 $1111$ systems RFeAsO
($R=$rare earth element), $122$ systems XFe$_2$As$_2$(X=alkaline earth
metals),   111 systems like LiFeAs.
  Examples of Fe-chalcogenides are
 FeSe, FeTe,
 and A$_x$Fe$_{2-y}$Se$_2$ ($A = K, Rb, Cs$).  The crystallographic structures of various families of FeSCs  is shown in Fig. \ref{fig:el_structure}.
 All FeSCs contain planes made of
 Fe atoms, with pnictogen/chalcogen atoms  above and below the iron planes.

\begin{figure}[tbp]
\includegraphics[angle=0,width=0.8\linewidth]{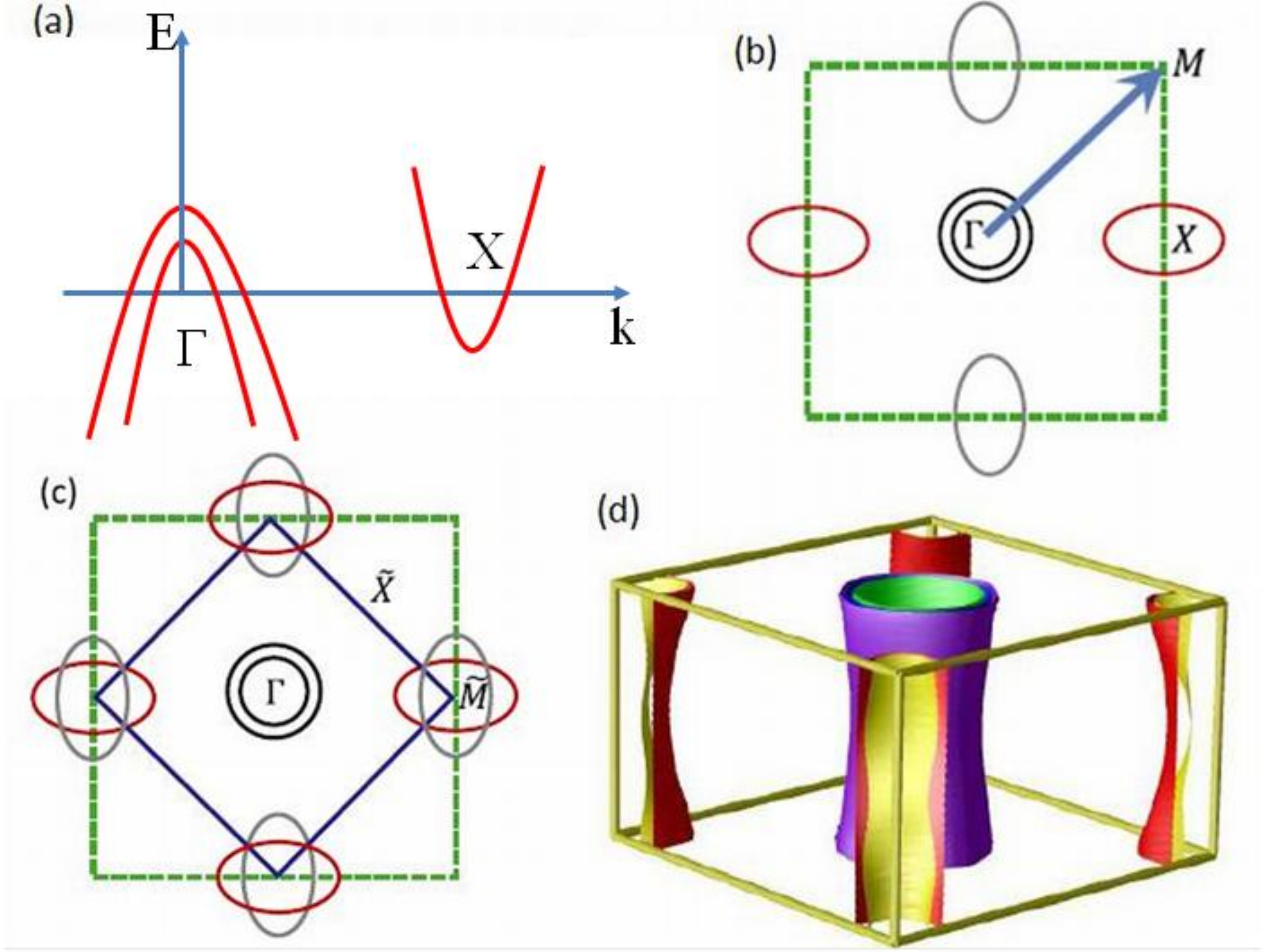}
\caption{The electronic structure of FeSCs. a) simplest schematic electronic structure E(k), with one hole-like and one electron-like pocket;  b) schematic 2D Fermi surface $E(k)=E_F$ for $k_z$=0 represented in 1-Fe zone;  c) schematic 2D Fermi surface in 2-Fe zone; d) full 3D Fermi surface for LaFeAsO calculated in density functional theory.  (Ref. \cite{mazin_schmalian}).}
\label{fig:FS}
\end{figure}

 The electronic structures of FeSCs at low energies are rather well established by
  band-structure  calculations
   and has been confirmed by
angle-resolved photoelectron spectroscopy  (ARPES) and other
 measurements.  At least three Fe orbitals $d_{xy}$, $d_{yz}$, and $d_{xz}$ contribute to
 the states near the Fermi surface,
and the hopping between Fe sites occurs primarily via a pnictogen (chalcogen) ion.
 Most FeSC have energy bands that
  are
   hole-like near the FSs centered at $(0,0)$
  (filled states are outside a FS) and electron-like
 near the Fermi surfaces (FSs)
  centered at $(0,\pi)$ and $(\pi,0)$  (filled states are inside a FS),
   as shown in Fig. \ref{fig:FS} a) and b).
Because electron and hole FSs are well separated in momentum space, they are often called hole and electron pockets, and we will use this notation below.   Since there are two inequivalent Fe positions in a crystalline unit cell,
the Fermi surface shown in Fig.  \ref{fig:FS} (b)
should be more properly viewed
 in
  a "folded" representation in the smaller reciprocal space unit cell, as shown in Fig.  \ref{fig:FS} (c). Viewed in 3D the Fermi surface usually consists of several corrugated cylinders, as shown in Fig. \ref{fig:FS}(d).


\begin{figure}[tbp]
\includegraphics[angle=0,width=\linewidth]{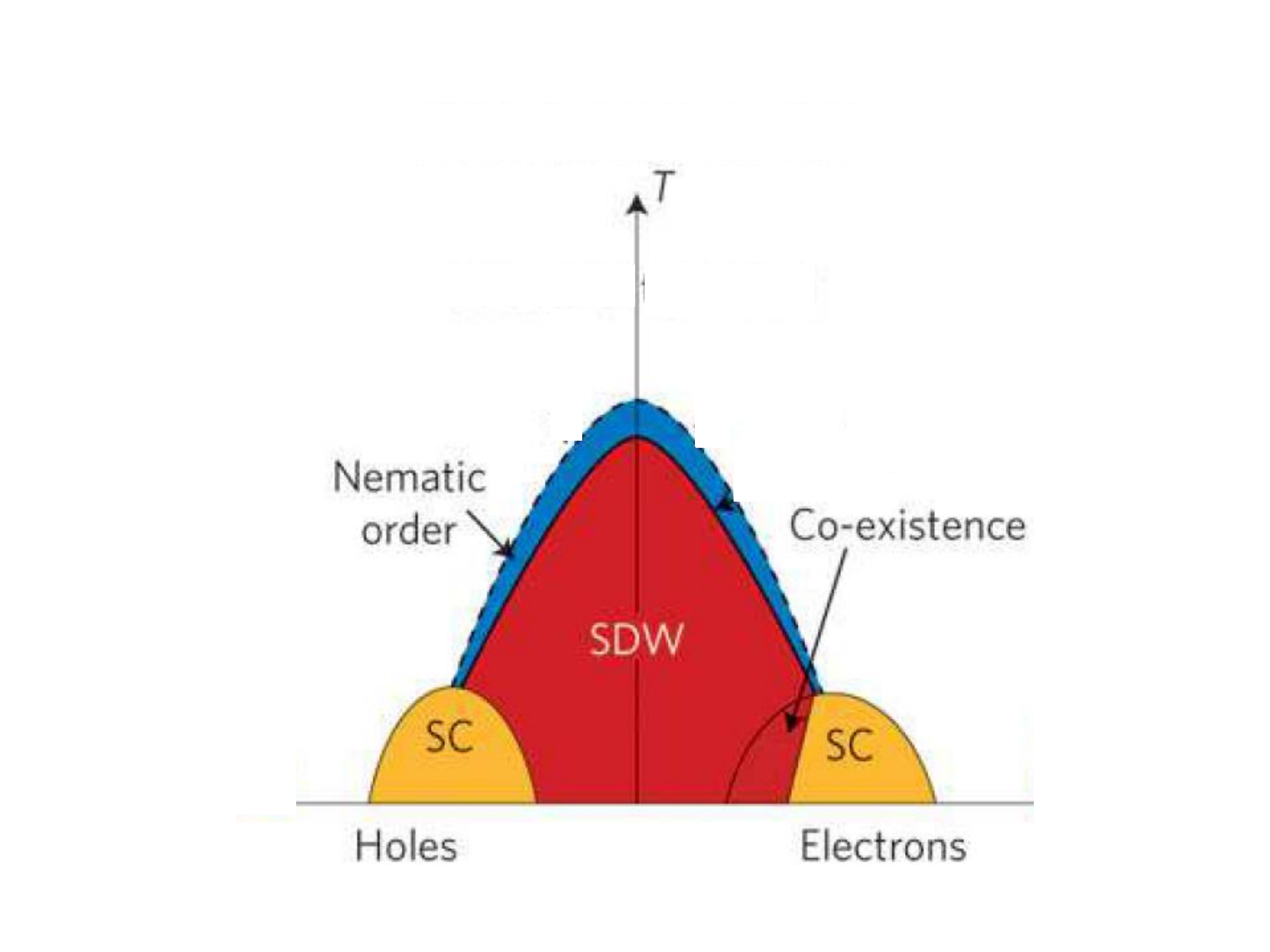}
\caption{{\bf Schematic phase diagram of Fe-based pnictides upon hole or electron doping.} In the red region, labeled SDW,
   the system has a magnetic order. In a yellow region, labeled SC, the system has superconducting order. In a blue region above SDW phase
     the system develops a nematic order.
     From Ref. \cite{andrey}a}
\label{fig1}
\end{figure}

The phase diagram of a "typical"  FeSC is shown in Fig. \ref{fig1}. The
 undoped (parent) compound is usually an  antiferromagnet
(with a few exceptions). The magnetic  phase of the FeSC is often called a spin-density-wave (SDW) to stress that this is magnetism of itinerant electrons rather than of localized electron spins. Upon  doping
   a parent compound,
   a superconductor is created.  This can be reached
   by substituting elements that add holes or electrons (e.g. replacing Fe by Co
    or Ba by K, which tips the balance of carriers in favor of electrons or holes, respectively),
    by applying pressure, or by
isovalent replacement of one  element by another (e.g., As by
 P).     There is also another
      ordered  phase termed  "nematic" by analogy to liquid crystals,  where the electronic state is believed to spontaneously break the symmetry between $X$ and $Y$ spatial directions without displaying magnetic or superconducting order.

\subsection{Magnetic  phase}

\begin{figure}
\includegraphics[angle=0,width=0.8\columnwidth]{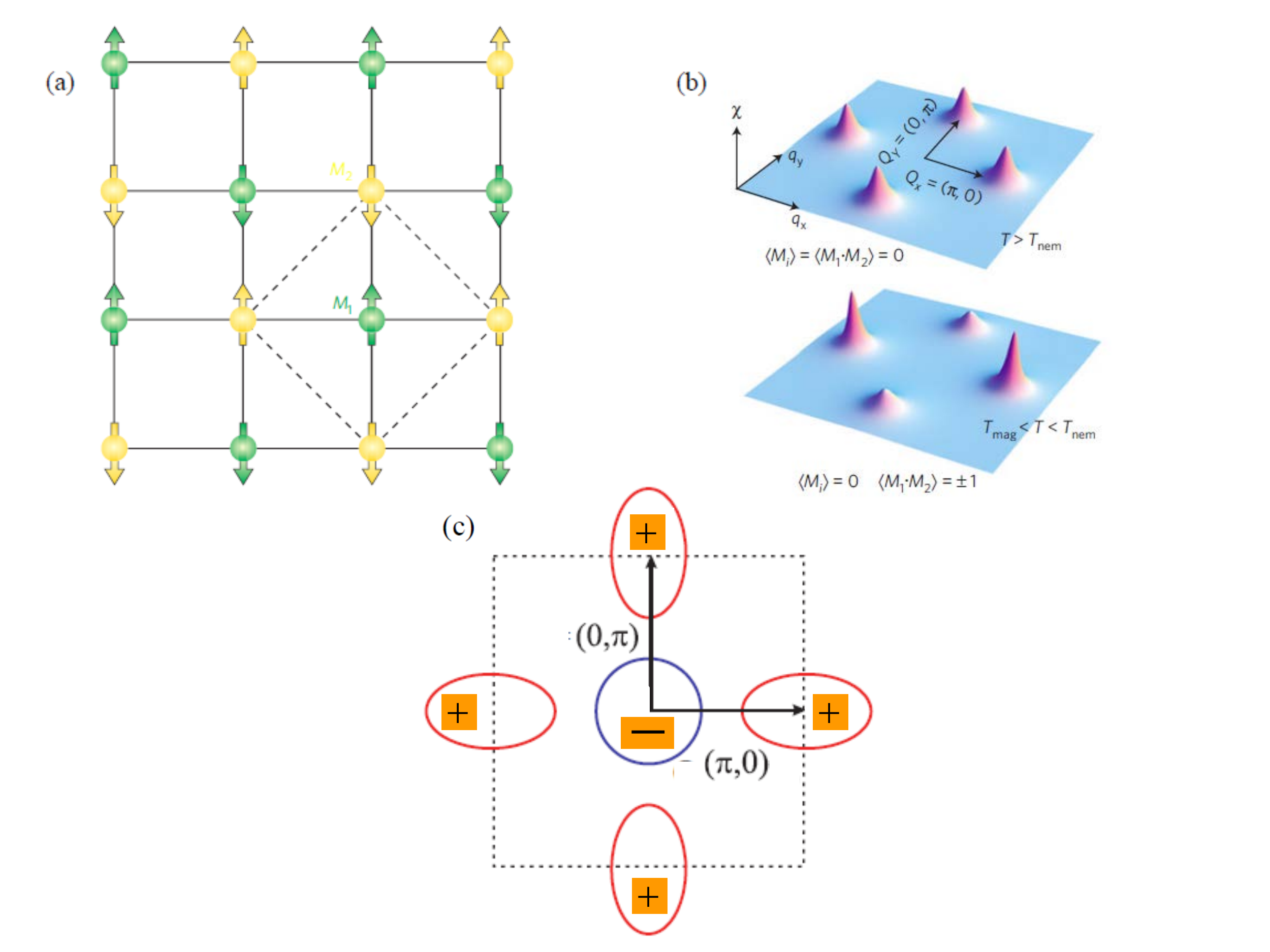}
\caption{\textbf{Magnetic, nematic, and superconducting order in Fe-pnictides} (a) -- stripe magnetic order at $T <T_{\mathrm{mag}}$.
This order is SDW with momentum $(0,\pi)$ (as shown) or $(\pi,0)$, but not both. A stripe order can be
interpreted as two inter-penetrating Neel sublattices (green and yellow)
with staggered magnetizations $\mathbf{M}_{1}$ and $\mathbf{M}_{2}$.
(b) -- nematic  order probed by magnetic susceptibility.
 For $T>T_{\mathrm{nem}}$, the two inelastic
peaks at $(\pi,0)$ and $(0,\pi)$
have equal amplitudes, i.e. $\langle \mathbf{M}_{1}\cdot\mathbf{M}_{2}\rangle =0$.
For $T_{\mathrm{mag}}<T<T_{\mathrm{nem}}$, one of the peaks becomes
stronger than the other, i.e. $\left\langle M_{X}^{2}-M_{Y}^{2}\right\rangle \equiv\left\langle \mathbf{M}_{1}\cdot\mathbf{M}_{2}\right\rangle \neq0$,
which breaks the equivalence between the $x$ and $y$ directions. (c) --  $s_\pm$ superconductivity due to repulsive interaction between hole and electron pockets, separated by the same momenta $(0,\pi)$ and $(\pi,0)$, at which SDW order develops.
 From Ref.\cite{rafael}.}
 \label{fig_new}
\end{figure}

This part of the phase diagram of FeSCs is best understood and least controversial.
Experiments have found
  that a magnetic order in {\it most}  undoped
  and weakly doped FeSCs
  is best described as
 stripe order, with spins ordering ferromagnetically in one direction and antiferromagnetically in the other direction in real space.
  (see Fig. \ref{fig_new} (a)).
 Such an order not only breaks $O(3)$ spin symmetry but
  also  an additional  $Z_2$ symmetry, as
  the stripes align either along $X$ or along $Y$.     Spin-orbit coupling requires that the lattice symmetry is simultaneously reduced from $C_4$ (tetragonal) to $C_2$ (orthorhombic).
 In some doped systems a small phase of  magnetic order
  preserving
   $C_4$ lattice symmetry has been discovered.

 Both $C_4-$breaking and $C_4$-preserving magnetic orders are consistent with the analysis of itinerant magnetism~\cite{andrey_2},
  where spin correlations building up
at a wave vector
 ${\bf Q}$ in a metallic system can drive a  transition to a SDW.    In Cr metal, it has been known for
 some time that this tendency can be enhanced by the presence of hole and electron pockets,  and this picture appears to hold for the FeSC,
  where ${\bf Q}$ connects the $\Gamma$- and $X,Y$-centered pockets in Fig. \ref{fig:FS}.
 A stripe magnetic order has also been
  obtained in the
  localized spin approximation~\cite{localized},
  in which one formally considers electrons
  as localized.
   There is still some outgoing debate about details (e.g., the form of magnetic excitations at energies of a few hundred meV), but, overall, the magnetically ordered phase is quite well understood.

 \subsection{Nematic phase}

Measurements of lattice parameters, dc resistivity, optical conductivity, magnetic susceptibility
 and other probes have found that  the stripe SDW order
is often preceded by a phase with broken $C_{4}$
tetragonal symmetry but unbroken $O(3)$ spin rotational symmetry (see Fig. \ref{fig_new}b).
   Such a state has been called a "nematic",
  by analogy with liquid crystals, to
 emphasize that the nematic order  breaks rotational symmetry but preserves time-reversal and translational symmetry.
 The debate about the origin of this phase has been extremely lively, since there are several possibilities:
 (i) a
   conventional
   structural transition caused by phonons, ; (ii) a spontaneous orbital order (specifically a difference in the occupation of $d_{xz}$ and $d_{yz}$ orbitals); and (iii) a splitting of a magnetic transition into a stripe SDW phase into two transitions with an intermediate partially-ordered Ising-nematic  state which breaks $C_4$ symmetry
 but does not break $O(3)$ spin-rotation symmetry.
 The last two scenarios attracted a lot of interest as they identify a nematic order as a spontaneous electronic order due to interactions.
   It is important to realize,
     however,
     that structural order, orbital order, and  Ising-nematic spin order break the same $C_4$  symmetry, hence the corresponding order parameters  are linearly coupled.
  A spontaneous creation of one
   then
   triggers the appearance of the other two,  leading to the subtle question of which phenomenon drives the others, i.e., which susceptibility actually diverges on its own upon approaching the nematic transition.
 In this respect, the strong measured enhancement of the resistivity anisotropy by strain
  seems to  argue against
     a
     structural transition (i)
      and favor a spontaneous electronic order scenario.
The observation that the SDW and nematic transition lines
follow each other across all the phase diagrams of 1111, 122
  materials,
 even inside the superconducting dome, has been suggested as evidence for the magnetic scenario~\cite{rafael}.
    On the other hand, in some systems like FeSe,  nematic order emerges when magnetic correlations are still weak, which has fueled speculations
    that at least in this system
  nematicity may be due to spontaneous orbital order.

 \subsection{Superconducting phase}

No matter how interesting the normal state is, the origin of superconductivity is always the primary objective, and understanding
what causes the pairing of electrons into Cooper pairs is the biggest goal in the studies of FeSCs.
Experimentally, superconductivity with $T_c$ up to
  nearly $60K$ has been detected in 1111 systems with both hole and electron pockets
The optimal superconducting temperature is somewhat smaller in 122 materials and varies between different compositions. In electron-doped
BaFe$_{1-x}$Co$_x$Fe$_2$As$_2$ superconductivity was found to disappear near the doping at which hole pockets vanish.
 In hole-doped
 Ba$_{1-x}$K$_x$Fe$_2$As$_2$, however,
 superconductivity is suppressed  but survives out  to a doping of $x=1$, beyond the concentration $x$ where electron pockets disappear.  In addition, superconductivity with rather high $T_c \sim 40K$ has been  found in
 A$_x$Fe$_{1-y}$Se$_2$ (A = K, Rb, Cs), which, according to ARPES,
  appears to contain only electron pockets, and even larger $T_c$ has been found in thin films of FeSe, which
  have a similar electronic structure.
\begin{figure}
\includegraphics[angle=0,width=0.8\columnwidth]{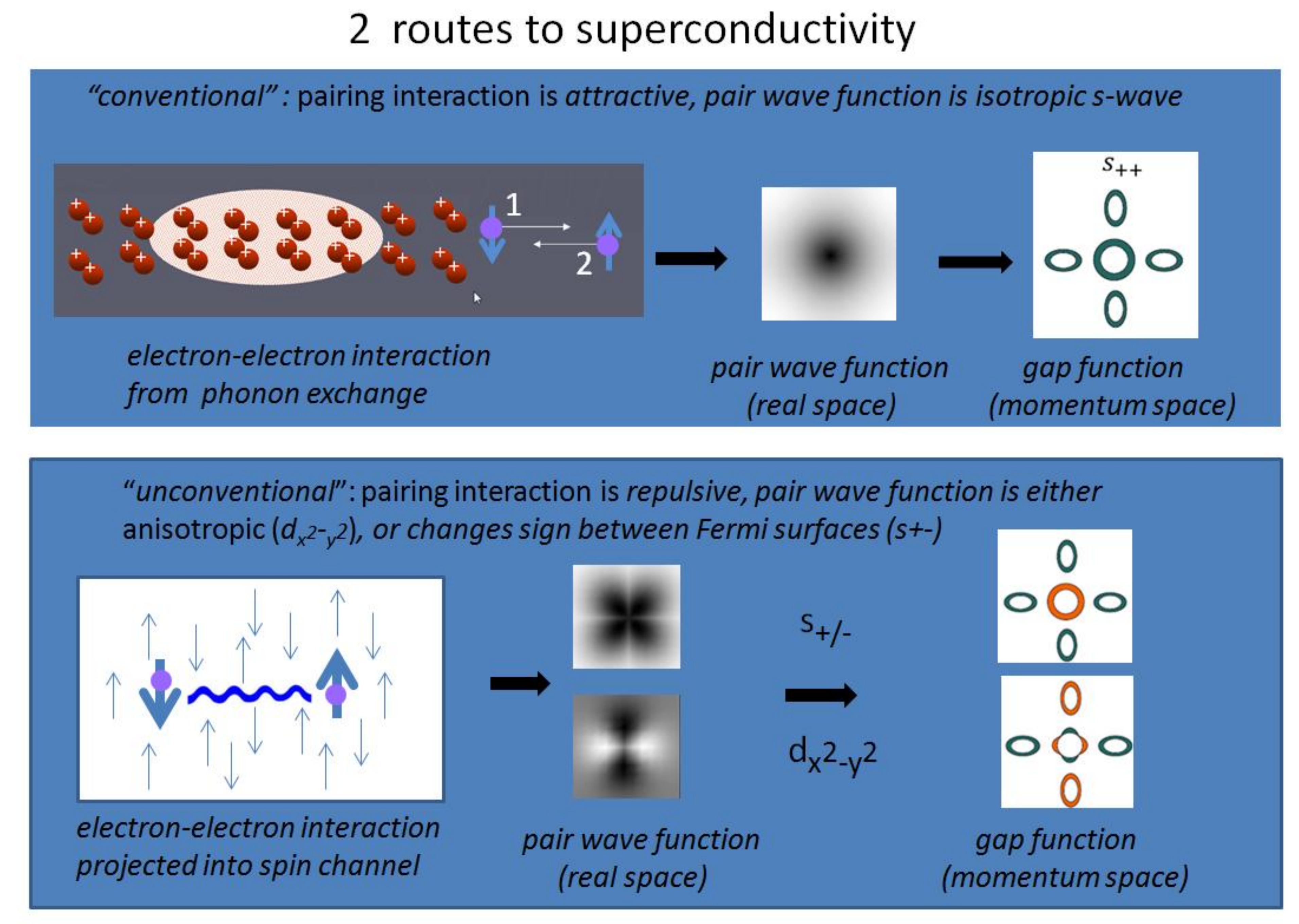}
\caption{\textbf{Two routes to superconductivity} (a) --
  two electrons attract each other when the 1st polarizes the lattice, and the second is attracted to this region.  The {\it pair wave function} $\psi$(${\bf r}$) of the relative electronic coordinate ${\bf r}$,   has the full symmetry of the crystal and gives rise to a {\it  gap function} of the same sign everywhere on the FeSC Fermi surface (green=+).
  (b) electrons  interact with each other via Coulomb interaction.   Shown is an example where the dominant interaction is the magnetic exchange arising between opposite spin
  electrons due to Coulomb forces.   The first electron polarizes the conduction electron gas antiferromagnetically,
  and an opposite spin electron can lower its energy in this locally polarized region.  In this case $\psi$(${\bf r}$) has a node at the origin, helping to avoid the Coulomb interaction, and can have either $s_{+/-}$ or $d_{x^2-y^2}$ form, as shown.
  These two possibilities lead to  gap functions of opposite sign on the Fermi surface (orange = -).}
 \label{fig:PHnew}
\end{figure}

From the  perspective of theory, the  central issue here is what  causes the attraction between electrons.
  In the
   Bardeen-Cooper-Schrieffer (BCS) theory of superconductivity,
    which was successfully used to describe many "conventional" superconductors,
  the two electrons effectively
  attract
  each other  by emitting and absorbing a phonon (a quantum of lattice vibrations) --
   one electron polarizes a lattice of positively charge ions and second electron is attracted into the same area
by the momentary accumulation of positive charge. This second electron, however, must wait a certain time until the first electron is out of the way to avoid the Coulomb interaction between the two electrons (for this reason, electron-phonon interaction is referred to as "retarded" in time).
  The phonon-mediated attraction binds fermions into a relative $s$-wave pair state and gives rise to an isotropic, roughly constant energy gap on the Fermi surface.  A schematic version of this ``conventional" scenario is depicted in Fig. \ref{fig:PHnew}a).
  For FeSCs,
  first-principles studies
  of superconductivity due to the electron-phonon interaction placed $T_c$ at around $1K$, much smaller that the actual $T_c$ in most FeSCs.  This  leaves a nominally repulsive  screened Coulomb interaction
   (that the Cooper pairs in conventional superconductors try to escape by being in the same place at different times)
  as the  most likely source of the pairing
  and puts FeSCs
  into the class of materials with electronically-driven superconductivity, like high $T_c$ cuprates.  The "unconventional" scenario for superconductivity is sketched roughly in Fig. \ref{fig:PHnew}b; it leads to highly anisotropic pair wave functions, and gap functions that change sign on the Fermi surface.

  The possibility of  superconductivity from electron-electron interactions is based on
  two fundamental principles,
  originally discovered for isotropic systems
    (see Insert I).
   In short,
        one can create superconductivity
        from
        non-zero
         angular momentum components of the
          screened Coulomb interaction,
          as they
          depend on the
          position of interacting electrons along the FS
         rather than on the overall sign of the
          interaction.
            The extension of these principles to FeSCs,
     for  which
      angular momentum  is no longer a good quantum number because FeSCs are
            crystalline systems with multiple FSs,
            implies that $T_c$ is non-zero if at least some
     interpocket interactions
      exceed intrapocket interactions.~\cite{peter,andrey}
      In most calculations done so far, both  intrapocket and interpocket interactions appear to be repulsive.
       In this situation,
       to convert repulsion into attraction,
       the phase of a $U(1)$ superconducting order parameter
        must change
         by $\pi$
        between pockets (see Fig.\ref{fig_new}c
        and Fig. \ref{fig:symmetry}b).    Such a state,  called $s^{+-}$ , is the analogue in multiband
         crystalline
         systems of the higher angular momentum pairings in isotropic single band systems
           and of $d_{x^2-y^2}$ superconductivity in the cuprates,
            but has the full symmetry of the crystal lattice (see
             Fig. \ref{fig:PHnew} and Insert II).
            In this respect, FeSCs provide the first example of {\it electronically-driven  $s-$wave superconductivity}.

  The reasoning for $s^{+-}$ superconductivity may look quite straightforward, but there is one
  major obstacle --
   the usual screened Coulomb interaction is larger at small momentum transfer (i.e., within one pocket) than at
    at larger momenta, connecting hole and electron pockets. To get $s^{+-}$ superconductivity,  one has therefore to invoke some mechanism to enhance interpocket interactions.
    The most popular scenario is that this is due to spin fluctuations (Fig. \ref{fig:PHnew}b)  which boost interpocket pairing
      because the  momentum connecting hole and electron pocket
     is the same ${\bf Q}$ as in the SDW-orderd state. Some groups working along these lines assume that magnetism is "superior" to superconductivity in the sense that
     spin fluctuations develop well above $T_c$.  This goes under the name
 "spin-fluctuation" scenario,
  as Charles Day explained in his  2009 review. Others do not
 assume a priori that spin fluctuations develop above $T_c$, but rather consider superconductivity and magnetism on an equal footing,
  and analyze how interactions in  magnetic and superconducting channels evolve as one progressively integrates out electrons
   with higher energies~\cite{rg}. This goes under the name "renormalization group" (RG) scenario.
    Several RG-based computation schemes have been proposed
    and they all lead to the same result as in the spin-fluctuation
   approach:  as magnetic fluctuations grow, they progressively increase the interpocket interaction which
     eventually
     becomes larger than  the intrapocket one
   and gives rise to an attraction in the $s^{+-}$ channel.

 The arguments for $s^{+-}$ superconductivity have been around since the early days after the discovery of Fe-based superconductors~\cite{mazin_schmalian,peter,andrey}. The majority of researchers in this field  believe that $s^{+-}$ is the right symmetry, even though
 the structure of  $s^{+-}$ superconducting order parameter has turned out to be more complex than originally thought (Insert II).
So far, however,
 there  has been no ``smoking gun" experiment which would settle this issue.
The experimental case  for $s^{+-}$ symmetry comes in two steps~\cite{review_ex}. First, there is evidence that the pairing symmetry in at least
 weakly and moderately doped FeSCs is $s$-wave. The most significant evidence comes from angle-resolved photoelectron spectroscopy  (ARPES) experiments, which show that
  $\Delta ({\bf k})$ on the central hole pockets
  does not have zeros on the Fermi surface
   except, possibly, some values of $k_z$.
   Only $s-$wave is consistent with this observation.  Secondly, there are several pieces of indirect evidence for $\pi$ changes of the phase of $\Delta (k)$ between hole and electron pockets.
    The most often cited evidence
  comes from neutron scattering experiments, which detected~\cite{Cruz} a resonance-like peak in the spin response
   below $T_c$  at momentum $(\pi,\pi)$.  If this peak is a true resonance, its existence implies that $\Delta (k)$ and $\Delta (k + (\pi,\pi))$ have opposite signs, similarly to the cuprates where the resonance peak was interpreted as strong, albeit indirect, evidence for $d-$wave symmetry.

What about the alternatives for the gap symmetry?  Two other states were
proposed for at least some FeSCs.  One is a conventional $s-$wave in which
 $\Delta (k)$ does not change sign between hole and electron pockets~\cite{kontani} (Fig. \ref{fig:symmetry}a).  A conventional $s-$wave superconductivity may be due to phonons, but it also emerges in the electronic scenario if the interpocket interaction dominates over the intrapocket one {\it and}
 is attractive. In multi-orbital systems like FeSCs,  the sign of interpocket interaction is determined by the interplay of various interactions
 in a basis of electron orbitals and can be attractive for some system parameters.
  An attractive interpocket interaction still needs an enhancement to overcome initially larger intrapocket repulsion, and some theorists say that this enhancement is provided by orbital fluctuations, much as
  a repulsive interpocket interaction
     is enhanced by spin fluctuations (an extension of these ideas to nematic order naturally leads to the prediction of spontaneous orbital order as the mechanism for nematicity).  A conventional $s-$wave state is inconsistent with the interpretation of a neutron peak as a resonance and is less likely for by this reason. However, in the absence of a true smoking gun experiment that probes the order parameter phase directly, $s^{++}$ superconductivity cannot be completely ruled out.

 Another alternative is $d_{x^2-y^2}$ superconductivity.  Numerical studies of the pairing
 in various channels show that
  the interaction in $d_{x^2-y^2}$ channel is attractive and is sometimes comparable in strength to the one in $s^{+-}$ channel.
      One rationale for  $d-$wave
        pairing
        comes from the consideration of a repulsive interaction between the two electron pockets
     If this interaction is somehow enhanced and exceeds
       other interactions, we again obtain a "plus-minus" superconductivity, but this time the sign change is between the gaps on the two electron pockets~\cite{peter}.
   By symmetry, this is a $d_{x^2-y^2}$ state,  since the superconducting order parameter $\Delta(k)$ changes sign under rotation from X to Y direction in the momentum space.
  In weakly and moderately doped FeSCs $d-$wave superconductivity  comes as close second behind $s^{+-}$, but it  emerges as the leading superconducting
   instability in  strongly electron-doped FeSCs, for which the electron-hole interaction is reduced.  The observation of a  change of  pairing symmetry in the same material upon doping would be unprecedented and is another reason why researchers are so excited about FeSCs.  Several groups have argued that, if the change of the pairing symmetry with doping  really happens, there must be an intermediate doping regime where superconductivity has $s +i d$ symmetry~(Ref. \cite{rg}b), i.e., both $s^{+-}$ and $d_{x^2-y^2}$ order parameters are present and the relative phase between the two is $\pm \pi/2$.  Such a superconducting state breaks time reversal symmetry and $C_4$ lattice rotational symmetry($s+id$ and $s-id$ are {\it different} states)
   and
   exhibits a wealth of fascinating properties like circulating supercurrents near impurity sites.
    $d-$wave superconductivity has also been proposed  (for different reasons) for strongly hole-doped FeSCs~\cite{louis} and is the subject of an intensive current research.

\section{Pnictides vs cuprates}

One of the main  sources of initial excitement surrounding the Fe-based superconductors was the hope that comparison to the cuprates
 might lead to a better understanding of the essential ingredients of high-$T_c$ superconductivity.
   The cuprate superconductors were discovered by Bednorz and Mueller in 1986 and hold the current record for $T_c$  at
    over 150K under applied pressure.
   The
   proximity
   of superconbducting region to an antiferromagnetically ordered phase in both classes of materials
     supported early suggestions that magnetic excitations might mediate superconductivity in both cases.   On the other hand,
  the  parent compounds of FeSCs
  with 6 d-electrons per Fe ion
  are
      metals, whereas the parent compounds of the cuprates
      with 9 d-electrons per Cu ion
       are invariably
 Mott insulators, in which electronic states are localized by strong
  Coulomb interactions.
    Evidence for Mott physics in FeSCs
   was unclear after the initial discoveries, and many researchers felt that good
    qualitative agreement of band structure calculations~\cite{band} with  ARPES experiments indicated that these systems were characterized by overall moderate electron-electron interactions,
     which are capable of giving rise to SDW magnetism and superconductivity at elevated temperatures, but not strong enough to localize the electrons.

Two discoveries have suggested that an understanding of larger class of Fe-based superconductors may require going beyond this scenario if the goal is to understand the system behavior
  over a wide range of energies.
  The first is that
   density functional theory  calculations are
   consistently found to give
   more dispersive bands than the measured ones.
     Second,  researchers have now created and studied Fe-based materials over
      a wide doping range, from close to $n = 5.5$ d-electrons per
       Fe  ion (e.g. strongly hole-doped KFe$_2$As$_2$) to
     $n \leq 7$ d-electrons in  strongly electron-doped systems, like  Ba(Fe$_{1-x}$Co$_x$)$_2$As$_2$.
      Specific heat data consistently show that mass renormalization, which roughly measures the strength of electron-electron interaction, grows as
       $n$
      decreases towards 5, corresponding to a half-filled $d$-shell.

Whether this last trend indicates the trend towards a Mott insulator, similarly to the cuprates, is the subject of debate.
 Some researchers argue
 that FeSCc are qualitatively different from the cuprates in the sense that the interaction effects are, at least to a certain extent, due to
 exchange (Hund) part of the Coulomb interaction~\cite{kotliar}. Strong Hund interaction destroys fermionic coherence but does not leads to insulating behavior, and the term Hund metals was introduced to describe systems with strong Hund interaction.  Others argue~\cite{luca} that FeSCs do have sizable density-density (Hubbard) interaction $U$, and the critical $U$ for Mott physics gets smaller as $n$
  gets closer  to 5.  An interesting new feature brought about by this last line of reasoning is the phenomenon termed as
 “orbital Mott selectivity”, which implies that critical $U$ is different for different orbitals, and some orbitals show stronger tendency to localization than the others as $n \to 5$.

 There is another issue which invites comparisons of FeSCs and the cuprates. In both sets of materials, resistivity shows a prominent linear in $T$
  behavior above $T_c$ near optimal doping (where $T_c$ is the highest).  There is no theory yet for linear in $T$ resistivity in a clean system down to $T=0$.
   However, there have been numerous attempts to link $\rho \propto T$ in the cuprates to fluctuation effects associated with a  putative quantum critical point --
   the end point of a possible phase transition line hidden by superconductivity.  In this respect, FeSCs
   may provide a simpler example, as the only two
quantum critical points currently known in FeSCs are associated with either SDW magnetism or nematicity.  Several groups are exploring the idea that fluctuations associated with one of these critical point may finally reveal the origin of linear in $T$ resistivity in
    FeSCs and
      then, possibly, in the cuprates and
    in other systems like heavy fermion materials as well.

 \section{New systems, new paradigms?}

  The paradigm established for the near-optimal 1111, 122 and 111 materials -- with $s_\pm$ pairing between central hole and outer electron pockets due to repulsive interpocket interactions -- has recently been challenged in some ``outlying" materials classes. The FeSC are famously more variegated than their cuprate cousins, so it is sometimes not so easy  to decide if these outliers with unusual properties represent a true challenge to this  paradigm or not.
 These "outlying" materials are mostly (but not exclusively) the
systems with large hole or electron doping.
Not only do these systems show the largest possible deviations from 6 d-electrons per Fe atom of parent compounds, but
 the low-energy electronic structure in these materials is quite different from that in weakly/moderately doped FeSCs.
 In systems with strong hole doping, like  KFe$_2$As$_2$, the electron band moves above Fermi level and only hole pockets remain.
 In systems with strong electron doping, like  A$_x$Fe$_{2-y}$Se$_2$ (A = K, Rb, Cs) most of ARPES data show that the opposite happens -- hole bands move below
   the Fermi level and only electron pockets remain. In both cases, one of two types of carriers which were apparently necessary for $s^{+-}$ superconductivity  disappears.
 Since superconductivity normally involves
  electrons
 near the FS, one might expect $T_c$ to  disappear
  or at least strongly decrease, if one type of
 FS pocket is removed.
  Yet this
  happens neither
   when hole pockets are removed, as evidenced
  by the cases of K$_x$Fe$_{2-y}$Se$_2$, where $T_c \geq 30K$;
 nor in  monolayer FeSe grown on strontium titanate substrates, where $T_c \sim 60K$ or even higher (see Insert III).
   One possibility is that the interaction between the two electron pockets is strong enough to produce superconductivity without hole pockets.
   As mentioned above, in this situation the pairing symmetry should be $d-$wave, although  more exotic $s^{+-}$-like states due to hybridization between the
    pockets and a conventional $s$-wave in case interpocket interaction is attractive, have been also proposed (see Insert II).
     Both scenarios fall outside of the standard paradigm for $s^{+-}$ superconductivity and show
    and provide another example of the richness of the physics of FeSCs.

     For strongly hole-doped KFe$_2$As$_2$, $T_c \sim 3K$ is small and still may be due to the interaction between hole pockets and gapped electron states.
      But several other possibilities were also put forward, which yield either  $d-$wave superconductivity or a different $s^{+-}$ superconductivity due to interactions between electrons
       solely near hole pockets.
        If interaction between hole pockets
         truly
         causes superconductivity, we have another example of a pairing mechanism outside of the standard paradigm.

     Finally recent experimental results on FeSe have raised the question of whether this simplest of FeSCs is also an outlier.
        This material was  discovered back in 2008 but received less attention than other FeSCs in part because of its low transition temperature, about 8K, and in part due to persistent difficulties synthesizing pure single crystals.     Substantial progress  growing bulk FeSe crystals has recently been achieved  using cold vapor deposition, and these samples are now possibly the highest quality of all the Fe-based systems.  At first glance, several properties of this material seem at variance with the phenomenology of the FeSC developed for the pnictides.    The compound is not magnetic, and spin fluctuations appear to be present at low temperatures only, rather than above the structural transition.  Upon application of pressure, $T_c$ grows from $8K$ to nearly $40$
        and can also be enhanced to the 30-40K  range by intercalation, either by alkali atoms as discussed above, or by alkali-ammonia moledular complexes.  These materials are currently difficult to make in homogeneous form, but  are nevertheless very intriguing because of the extreme sensitivity of $T_c$
        (Insert III).       Whether the FeSe results require a new paradigm for
              iron-based superconductivity is currently being hotly debated.

\section{What's next for Fe-superconductors}

Perhaps the most amazing thing about the FeScs is the unprecedented
richness of the physics. Practically all phenomena associated with
strongly correlated electron
  systems  have been found in Fe-based materials, sometimes all within a single subfamily:  magnetism,
unconventional superconductivity, quantum-criticality, linear in $T$
resistivity,  nematic order,
   Hund metallicity, and a tendency towards orbital selective Mottness, to
name a few.  Besides this, FeSCs, with their multiple Fermi pockets,
 are the most likely candidates for the observation of
    a change in the pairing symmetry in the same material upon doping,
and therefore  also for the development of
 mixed superconducting
order which breaks time-reversal symmetry (e.g., $s+id$ or $s+is$). Such
states have a rich phenomenology and  strong potential for applications.

    Although a "smoking gun" proof is still lacking, it is likely that the
superconducting state in  weakly/moderately doped FeSCs has $s^{+-}$ symmetry,
and  magnetic fluctuations are the primary suspects to mediate this kind
of pairing. What happens at stronger hole and, particularly, electron
doping is an important open question,
 and the  high transition temperature found in
FeSe films which apparently only have electron pockets raises
the possibility that the pairing mechanism in this materials may represent a
completely  new paradigm for superconductivity in these materials.

       The number of FeSCs keeps growing, and there is  very high
probability that materials with higher $T_c$ and with qualitatively new
features will be found.
       But the volume of already existing experimental data is sufficient to
create enough puzzles for the community working on FeSCs and keep
the  level of excitement (and the intensity level of the discussions) quite high
for years to come.

\section{Inserts}

\subsection{Insert I: Superconductivity from repulsion -- isotropic systems.}

\begin{enumerate}
\item
The BCS equation for the critical temperature decouples into independent equations for each
 pairing channel characterized by its own
    angular momentum $l =0, 1,2,3$,
    [in spatially isotropic systems, the  $l=0$ component is called $s-$wave, $l=1$ component  $p-$wave, $l=2$ component $d-$wave, and so on]. If just one angular momentum component of the pairing interaction with some $l$ is attractive, the system undergoes a superconducting
     transition at some non-zero temperature $T=
     T^{(l)}_c$.
  \item
       The screened Coulomb interaction $U(r)$ is positive at short distances but oscillates at large distances.
        Kohn and Luttinger   explicitly demonstrated
         in 1965
        that this ``overscreening"  necessarily gives rise to
        attractive angular momentum components of the pairing interaction,   at least for large odd $l$.
\end{enumerate}

\subsection{Insert II: $s^{+-}$ state through a microscope}

\begin{figure}[tbp]
\includegraphics[angle=0,width=0.8\linewidth]{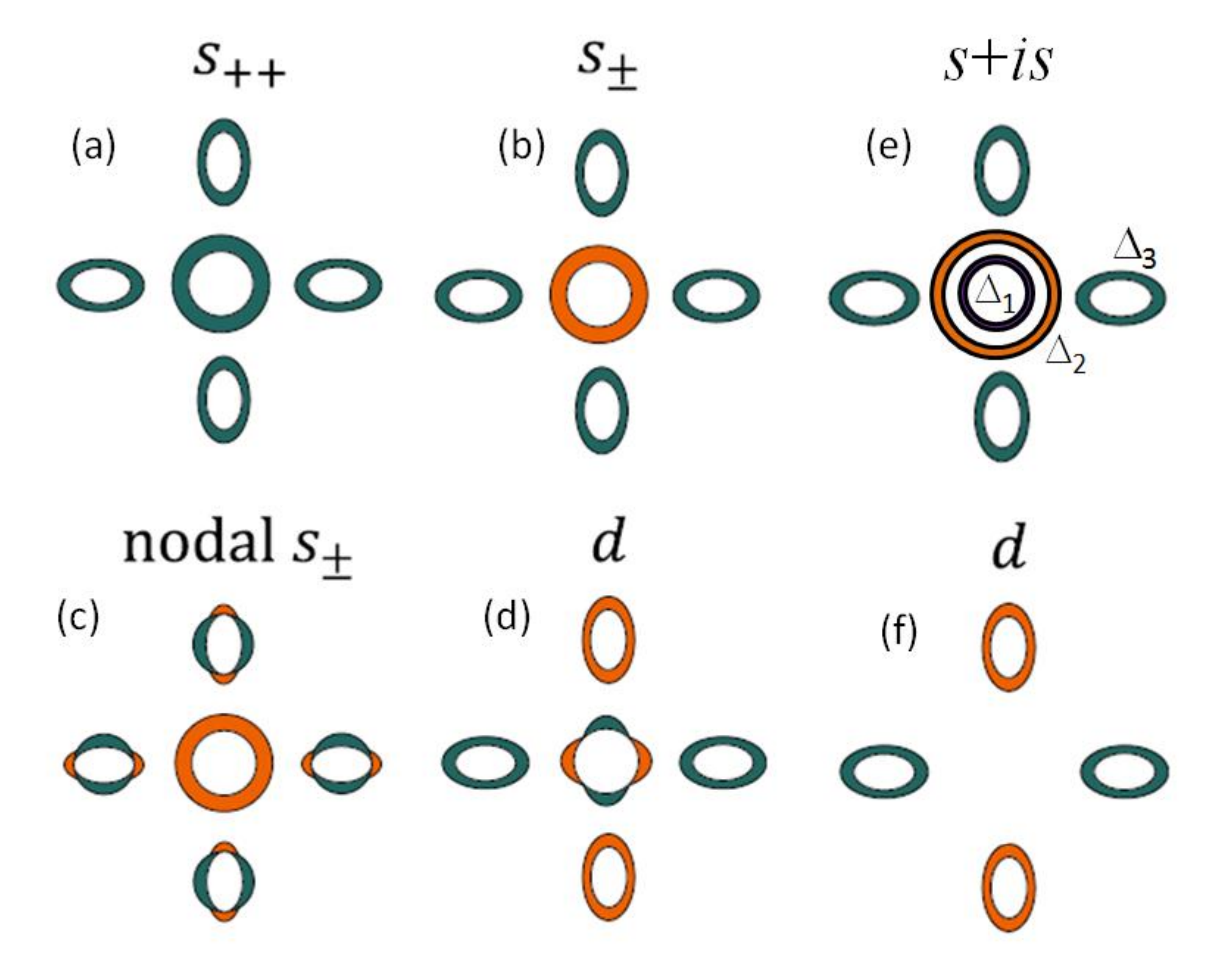}
\caption{Schematic gaps $\Delta(k)$ in FeSC.  Color represents phase of $\Delta(k)$.    (a)-(d) Model Fermi surface with   one hole and two electron pockets.  (a) Conventional $s$-wave ($s_{++}$) state; (b) $s_\pm$ state with gap on hole pocket minus that on electron pockets; (c) similar to (b), but with accidental nodes on electron pockets; (d) $d$-wave state.  (e) Possible state in situation with more than one hole pocket, showig gaps with three different phases $\Delta_i$. (f) $d$ wave state in situation with no central hole pocket.
    }
\label{fig:symmetry}
\end{figure}

 The structure of the $s^{+-}$ order parameter $\Delta (k)$ has
 turned out to be a subtle issue.
   In
   the simplest scenario, the gaps on hole and electron FSs
    are treated as constants and only differ in sign
    (Fig. \ref{fig:symmetry}(b)).  It was soon realized,
    however, that because of  the multi-orbital nature of FeSCs,  an $s^{+-}$ order parameter
    on each pocket necessarily has an angular variation which may be quite substantial. In particular, in the 1-Fe zone of Fig. \ref{fig:FS}, the angular variation of the order parameters on the two electron pockets is $\Delta (k) = \Delta_e (1 \pm \alpha \cos 2 \theta)$, where $\theta$ is the angle counted from $X$ direction (for both electron FSs).  If $|\alpha| >1$, $\Delta(k)$ has four nodes on each FS (Fig. \ref{fig:symmetry}(c)). These nodes have been called "accidental" as their position is
   not set by symmetry, as opposed to e.g., $d-$wave nodes (Fig. \ref{fig:symmetry}(d)) which by symmetry must be along certain directions in the Brillouin zone.   Note that if there is no central hole pocket, a $d$-wave state
   need {\it not} have nodes if the FS avoids these directions (Fig. \ref{fig:symmetry}(f)).  The presence or absence of the nodes is highly relevant, as it completely changes the low-temperature behavior of a system compared to a conventional $s-$wave superconductor.

An even more subtle issue is the actual structure of the phases between superconducting order parameters in a generalized $s^{+-}$ state.  We considered the case when the phase
 changes by $\pi$ between hole and electron pockets, but in multi-band systems other cases are possible, e.g., $s^{+-}$ gap between hole pockets, or phase differences which are not multiples of $\pi$ (Fig. \ref{fig:symmetry}(e)). In the last case, $s^{+-}$ superconducting order breaks time-reversal symmetry (it was termed $s+is$ for this reason).

\subsection{insert III - FeSe monolayers}

\begin{figure}[tbp]
\includegraphics[angle=0,width=\linewidth]{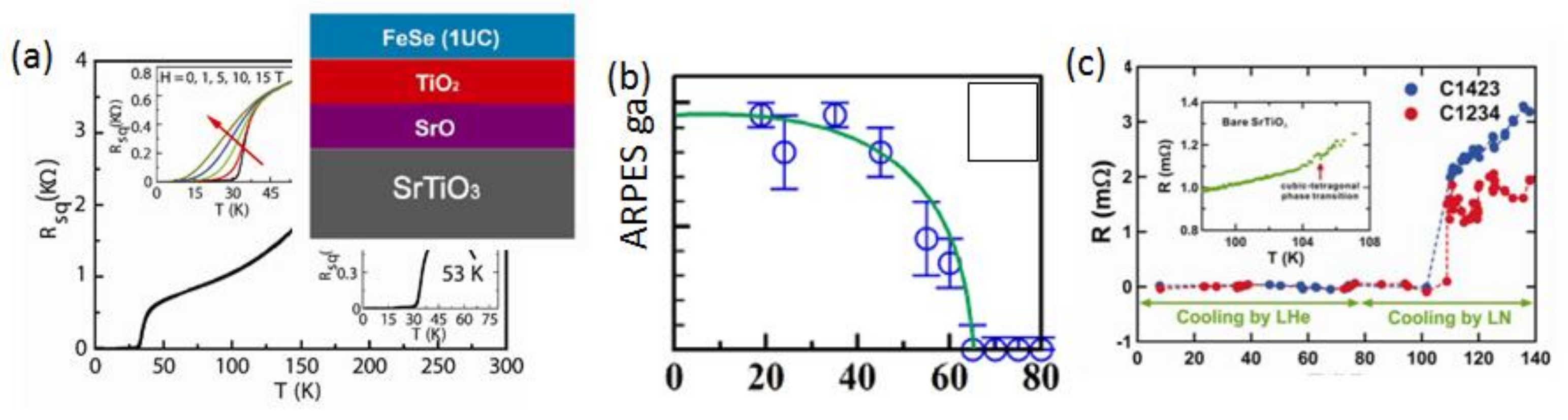}
\caption{(a) Resistivity of monolayer FeSe films on STO\cite{Xue}a;  (b) Spectral gap measured by ARPES on such films.  From Ref. \cite{Xue}b; (c) Resistivity of newer films of FeSe/STO\cite{Xue}c . }
\label{fig:FeSe}
\end{figure}

The most spectacular FeSe-based material has certainly been monolayer FeSe grown epitaxially on strontium titanate (STO),  by the Institute of Physics (Beijing) group led by X.-K. Xue in 2012\cite{Xue}.   This system was shown, after careful treatment of the substrate and annealing, to exhibit signs of superconductivity at very high temperatures  although, surprisingly, the 2-layer film grown by the same technique was not superconducting at all, indicating the importance of proximity of the active electronic layer to the substrate.
While zero resistance in these initial monolayer films was attained only below 35K (still much higher than the
8K bulk $T_c$), Fig.  \ref{fig:FeSe}(a),  the large gap measured in the electronic spectrum by ARPES vanished at a temperature of closer to 65K (Fig.  \ref{fig:FeSe}(b)).    Subsequent refinements have raised the ARPES gap closing temperature to 75K, not far from the symbolic temperature of 77K where nitrogen liquefies.
  The ARPES measurements indicate that the electronic structure of the monolayers resembles the alkali intercalated FeSe systems:
    the band normally responsible for the $\Gamma$ centered hole pocket is located many tens of meV below the Fermi level.

The high-temperature superconductivity in monolayer films and the ARPES results were confirmed recently by the Z.X. Shen group at Stanford.
   However in April came another, long-rumored surprise: when the Xue group performed in situ measurements of resistivity with a 4-probe "fork" pressed into the sample, they found~\cite{Xue}b that the resistivity disappeared below 108K  (Fig.  \ref{fig:FeSe}(c)).  If confirmed, this would be a clear record for the critical temperature of Fe-based systems.  Already, the result has inspired a number of theoretical suggestions, including ``bootstrapping" the superconductivity caused by repulsive Coulomb interactions by adding the binding forces due to exchange of phonons in the substrate and enhanced spin fluctuations due to the tensile strain to which the monolayer is subjected  by the STO.

\end{document}